\documentclass[aps,prb,twocolumn,superscriptaddress]{revtex4}
\usepackage{graphicx}
\usepackage{natbib}
\usepackage{subfigure}
\usepackage{tabularx}
\usepackage{epsfig}

\begin{document}

\title{Density, structure and dynamics of water: the effect of van der Waals interactions}

\author{Jue Wang}
\affiliation{Department of Physics and Astronomy, Stony Brook University, Stony Brook, New York 11794-3800, USA}
\author{G. Rom\'an-P\'erez}
\affiliation{Dep. de F\'{\i}sica de la Materia Condensada,
			Universidad Aut\'onoma de Madrid, 28049 Madrid, Spain}
\author{Jose M. Soler}
\affiliation{Dep. de F\'{\i}sica de la Materia Condensada,
			Universidad Aut\'onoma de Madrid, 28049 Madrid, Spain}
\author{Emilio Artacho}
\affiliation{Department of Earth Sciences,
             University of Cambridge,
             Downing street, Cambridge CB2 3EQ, UK}
\author{M.-V. Fern\'andez-Serra}
\email[To whom correspondence should be addressed:\\ ]{maria.fernandez-serra@stonybrook.edu}
\affiliation{Department of Physics and Astronomy,  Stony Brook University, Stony Brook, New York 11794-3800, USA}
\affiliation{ New York Center for Computational Science, Stony Brook University, Stony Brook, New York 11794-3800, USA}

\date{\today}

\begin{abstract}

	It is known that {\it ab initio} molecular dynamics (AIMD) simulations 
	of liquid water, based on the generalized gradient approximation (GGA)
	to density functional theory (DFT), yield structural 
	and diffusive properties in reasonable agreement with experiment only if
	artificially high temperatures are used in the simulations.
	The equilibrium density, at normal
	conditions, of DFT water has been recently shown by Schmidt {\it et al.} [J. Phys. chem. B, 113, 11959 (2009)] 
	to be underestimated by different GGA functionals for 
	exchange and correlation, and corrected by the addition of interatomic pair potentials
	to describe van der Waals (vdW) interactions.
	In this contribution we present a DFT-AIMD study of liquid water using several GGA functionals
	as well as the van der Waals density functional (vdW-DF) of Dion 
	et al. [Phys. Rev. Lett. \textbf{92},  246401(2004)]. 
	As expected, we find that the density of water is grossly underestimated by GGA functionals. 
	When a vdW-DF is used, the density improves drastically
	and the experimental diffusivity is reproduced without the need of thermal corrections.
	We analyze the origin of the density differences between all the functionals. We show that the vdW-DF increases the population of non-H-bonded interstitial 
	sites, at distances between the first and second coordination shells.
	However, it excessively weakens the H-bond network, collapsing the
	second coordination shell. 
	This structural problem is partially associated to the choice of
	GGA exchange in the vdW-DF. We show that a different choice for the exchange
	functional is enough to achieve an overall improvement both in structure and diffusivity.
\end{abstract}


\maketitle 
%Look at every possible position to add in the PBE 1.0 data point result
\section{Introduction}
Studies with density functional theory (DFT)-based
\textit{ab initio}  molecular dynamics (AIMD) simulations of liquid
 water\cite{Grossman04,Schwegler04,Asthagiri03,mvfs04,sit05,lee06,guidon08}
seriously differ from experimental results both in diffusive and structural properties.
This is true even though most studies have been performed at the experimental
density, what implies theoretical pressures as high as 1 GPa.\cite{Schmidt09}
The origin of the discrepancy with
experiments\cite{waterexpD,RDFexp1,RDFexp2} is still unclear.
One of the main conclusions of these studies\cite{mvfs04} was that
the AIMD results for radial distribution functions (RDF) and self-diffusivity at a given temperature 
compare well with the experimental results at a temperature 20\% lower.
Therefore, a practical solution to the problem is to perform 
simulations at temperatures 20\% higher than the desired reference
temperature,\cite{allesch08} as proposed in ref~\onlinecite{mvfs04}.
However, although this \textit{ad hoc} solutions may
be convenient to study other properties of liquid water,\cite{ManuPRL}
of molecules in solution,\cite{allesch08}
and of wet interfaces,\cite{cicero08} they do not help 
explain the origin of the differences between experiments and AIMD simulations. 

There are obvious limitations in the AIMD description of
liquid water that could account for these differences, including the inability of
present generalized-gradient approximation (GGA) density functionals to describe dispersion
interactions, or the complete neglect of quantum fluctuations
in the classical treatment of nuclear dynamics.
The latter question has already been addressed: Morrone and Car~\cite{morrone08} 
have found that when nuclear quantum effects are accounted for,
the simulated liquid becomes less structured and more diffusive, and the
agreement between AIMD simulations and experiments 
considerably improves. On the other hand, this result is at odds with
a previous similar study in which Chen \textit{et al}\cite{Chen03} 
found that the quantum vibrations of the hydrogen atoms would indeed
increase the strength of the hydrogen bonds in liquid water. Recently, 
Habershon \textit{et al}\cite{Manalopoulos09} identified the competing contributions 
from intermolecular and intramolecular quantum fluctuations to cancel out and thus 
to have a small net effect on the diffusion coefficient.  

Computational studies of liquid water differ widely in their methodology: e.g. they might employ either Born-Oppenheimer or Car-Parrinello\cite{CPMD} dynamics, they can use different types of basis sets,\cite{Grossman04,mvfs04,VondeVondele04,lee06} and nuclear quantum effects may or may not be accounted for. 
Nonetheless, most are performed at a fixed volume, set approximately to the experimental density of water at room temperature, $\simeq$1.0 g/cm$^3$. 
This approach would be correct if the calculated pressure also reproduced the experimental pressure at normal conditions (1 atm). 
However, as pointed out in Refs.~\onlinecite{mvfs04,siepmann}, even though pressure fluctuations are very large, due to the small size of the simulated system, the observed average pressure is much larger than the experimental one. 
Therefore, both structural and diffusive features correspond to a region of the temperature-pressure phase diagram which differs form the experimental values they are compared to.\cite{siepmann}
In a word, the simulations are performed at too high temperatures and pressures.

 Schwegler \textit{et al}\cite{Schwegler00} have already presented a study of DFT water under pressure, but the equilibrium density of GGA water
was not explored.
In Ref.~\onlinecite{siepmann}, the discrepancy between theory and experiments of the pressure-density phase diagram was already addressed using Monte Carlo simulations.
A recent study by Schmidt {\it et al.}\cite{Schmidt09} of liquid water
under isothermal and isobaric condition, using both GGA and so-called
 DFT+dispersion (DFT-D) method has analyzed this question in detail.
 They have shown that two of the most commonly used GGAs, BLYP\cite{BLYP1,BLYP2}
 and PBE\cite{PBE} have an equilibrium density of 0.75 and 0.88  g/cm$^3$  respectively, 25\% and 12\%
below the experimental value. 
When vdW interactions are added, using the interatomic pair potentials proposed by Grimme,\cite{Grimme}
the experimental density is recovered.

In this study, we partly confirm and extend their results by using AIMD but, more
importantly, we provide insights into the origin of the vdW effects: by
comparing results with the GGA and with the first-principles vdW density functional (vdW-DF) of Dion \textit{et al},\cite{DRSLL} and by separating H-bonded and non-H-bonded
interactions, we are able to explain the differences in density and local structure
between GGA water and vdW-DF water.
	
\section{Methodology}

%%

% Changes in this part, explaining why our functional choice.
The simulations are performed using the self-consistent Kohn-Sham approach \cite{KS} to DFT \cite{DFT} 
within the generalized-gradient approximation to exchange and correlation (XC). 
We choose two commonly used GGA functionals: PBE \cite{PBE} and revPBE.\cite{revPBE}
The choice of these functionals is motivated by our interest on studying 
the effect of dispersion interactions using the vdW-DF.\cite{DRSLL}
The original vdW-DF,  that we coin DRSLL after its authors,\cite{DRSLL} uses revPBE exchange combined with the non-local vdW correlation.
Since this choice was somewhat arbitrary, we have additionally substituted it by PBE exchange,
labeling this functional DRSLL-PBE. 
Besides addressing the effect of the exchange part in vdW-DF, this also provides a second reference point on the effect of non local correlations
on GGA functionals, i.\ e.\  DRSLL vs revPBE and DRSLL-PBE vs PBE.
We have used the implementation of vdW-DF by Roman-Perez and Soler\cite{SolervdW} 
within the {\sc siesta} program.

\subsection{Basis set}

Core electrons were replaced by norm-conserving pseudopotentials \cite{pseudo1} in their fully nonlocal representation.\cite{pseudo2}
Numerical atomic orbitals of finite support were used as basis set, and the calculation of the self-consistent Hamiltonian and overlap matrices 
was done using the {\sc siesta} method.\cite{siesta1,siesta2}
In this study, a double-$\zeta$ polarized (DZP)  basis set was used, 
which has been variationally optimized following the method proposed in Refs.~\onlinecite{basis1,basis2}.
The validation of the method, pseudopotentials and basis set can be found in Ref.~\onlinecite{mvfs04}. 
In addition, a triple-$\zeta$ polarized (TZP) basis set has also been tested in this study. The summary of the basis test result is presented
 in Table \ref{tab:basisTest}. We provide the details of these two sets in the Supplementary Information.
%Marivi, could you read this paragraph once to make sure what I am claiming here is right to include in this paper.

\begin{figure}[!hbtp]
\begin{center}
\includegraphics*[width=8.0 cm]{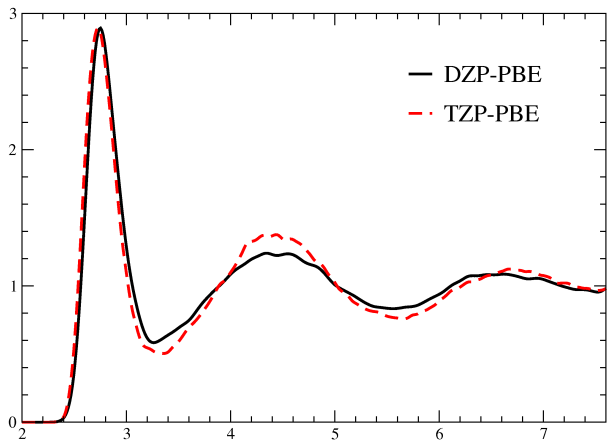}
\caption{Comparison of the O-O radial distribution function (RDF) obtained with the DZP(solid black) and TZP(dashed, red) basis sets (see text) at 1.0 g/cm$^3$. }
\label{GOO_basistest}
\end{center}
\end{figure}

Our results show that the main difference
between the DZP basis set we have used in this study and the well converged TZP basis is that the smaller basis produces
a somewhat less structured liquid. The diffusivity is approximately the same for the two basis sets. 
The first peak in the O-O radial distribution function (RDF) is almost identical, the first minimum is 13$\%$ higher. 
 The comparison of the radial distribution function between the TZP basis and the DZP basis is shown in Fig.~\ref{GOO_basistest}. The error of the DZP basis on the location of the first maximum
in the O-O RDF is less than 0.6$\%$, indicating that the basis is very good in the description
of the H-bond geometry, as already shown in Ref.~\onlinecite{mvfs04}. 
% Jue completely modified this paragraph on Aug 22nd, 2010

In order to evaluate total energy errors associated
to our choice of basis set, we have selected 10 snapshots, separated by 2 ps time intervals, from our simulation density 1.00 g/cm$^3$.
The difference between the TZP and DZP energies is an almost constant value of 0.12 eV/molecule, with a standard deviation of  only 7 meV, which can be neglected even when we are studying the small van der Waals effects. 
Also, in Table \ref{tab:basisTest}, the details of the two simulations with DZP and TZP are compared. The energy drift is on the order of 10$^{-6}$ a.u./atom/ps, which accounts for a change of only a few Kelvin during 20 ps, well below the statistical fluctuation of temperature from the MD simulation. We have estimated that basis-set-induced errors in the  equilibrium  pressure are on the same order of the error bar associated to the choice of different initial configurations, which is approximately 1 kbar. Both errors are much smaller than the statistical fluctuations of pressure during the MD simulation, which is on the order of 3-4 kbar.  Overall, we can conclude that AIMD of water are not totally insensitive to the basis set choice,\cite{mvfs04,VondeVondele04} 
shorter basis sets providing slightly less structured liquids in general. 
However, the results and conclusions presented in this study are well beyond the uncertainties due to these limitations.
% and they provide a new and meaningful insight to the problems of DFT in the description of liquid water. 

\begin{table}[h]
\centering
\caption{Summary of AIMD simulations results, for liquid water at 1.0 g/cm$^3$ and 340 K, using the PBE functional. We compare results with a DZP basis using a confining pressure of 0.2 GPa with a run of 20 ps, and a TZP basis using a confining pressure of 0.2 GPa with a run of 15 ps.
 Temperature $T$ (K), pressure $P$ (kbar), energy drift during the simulation $E_{drift}$ (10$^{-6}$ a.u./atom/ps), self diffusion coefficient $D$ (10$^{-5}$cm$^{2}$/s), height of first peak $g^{max}_{OO}$, height of first minimum $g^{min}_{OO}$ and position of first maximum
$r^{max}_{OO}$ (\AA) of the O-O radial distribution function  are compared.}
\begin{tabular}{c c c c c c c c}
\hline
\hline
Basis & $T$ & $P$  & $E_{drift}$ & $D$ & $g^{max}_{OO}$ & $g^{min}_{OO}$ & $r^{max}_{OO}$\\
\hline
%% Jue modified the table on Aug 20th, 2010
DZP & 353 & 4.5  & -1.5 & 1.23 & 2.87 & 0.58 & 2.75 \\
TZP & 365 & 5.8  & -2.3 & 1.20 & 2.85 & 0.50 & 2.73 \\
\hline
\hline
\end{tabular}
\label{tab:basisTest}
\end{table}

%Jue added at Aug 22nd, 2010

\subsection{Molecular dynamics simulations}

All the results are based on AIMD simulations of 64 molecules of heavy water. 
 Classical molecular dynamics simulations, where the electronic potential seen by the nuclei is replaced by an empirical force field, were performed before AIMD equilibration in order to prepare reasonably equilibrated initial configurations. 
 All empirical-potential-based simulations were performed with the TIP4P 
 force field\cite{TIP4P}  as implemented within the GROMACS MD package \cite{Gromacs1,Gromacs2} under constant volume and temperature conditions, with a Nose-Hoover thermostat \cite{Thermostat1,Thermostat2} at 300 K, along a 1 ns equilibration trajectory.
AIMD simulations were started from these pre-equilibrated systems imposing an additional 3 ps equilibration by means of temperature annealing (velocity rescaling).\cite{Annealing}
The actual production runs of the AIMD simulations (20 ps each) were accomplished by constant-energy
 Verlet's integration, given our interest in dynamical properties.
The time step in all simulations, (including empirical force field MD and AIMD) is 0.5 fs. 

The simulations are performed under constant volume i.\ e.\ fixed cell size and shape, 
under periodic boundary conditions. Rather than performing a 
constant-pressure simulation to establish the theoretical equilibrium density under ambient conditions,  we chose to perform a series of simulations by changing the volume of the unit cell while keeping the number of water molecules fixed. 
The reduced system size produces large pressure and temperature fluctuations and 
requires larger statistics for reliable comparisons between different simulations. 
Therefore, we compute three different trajectories for each density, starting from different equilibrations.
In total, over 500 ps of AIMD simulations were produced in this study. 

The temperature of the simulations using GGA functionals were set higher than 
300 K due to the reason mentioned in the introduction. 
With the PBE functional we use $\sim$360 K\cite{mvfs04} and 
with revPBE we use $\sim$330 K.\cite{mvfs04-1}

\subsection{System size effects}

System size effects on structural and diffusive properties of simulated liquid water have been addressed before.
Most studies\cite{Grossman04,RDFexp2} used empirical potentials and found that size effects seem not to be a problem for those properties.
In AIMD simulations, size effects have been shown to be rather small for structural properties,\cite{mvfs04,Parrinello09} but much larger for dynamical properties,\cite{Kremer9309,Parrinello09}meaning that finite size scaling is needed to obtain the infinite-size self-diffusivity coefficient.\cite{Dunweg93} 
With such finite size scaling, Kuhne \textit{et al} found\cite{Parrinello09} an improved self-diffusivity but still smaller than experiment.

We have explored the influence of size-dependent boundary conditions on both structural and dynamical properties of the system. 
The dependence of average pressure on the system size has also been studied in detail. These 
analyses were performed by means of classical molecular dynamics simulations. 
It seems reasonable to assume that size effects are robust enough to the change of interaction potential and that results obtained from force-field-based simulations are adequate to estimate the error due to the reduced size of the system in our AIMD studies. 

We have performed molecular 
dynamics simulations of five different system sizes (32, 64, 128, 512 and 1024 molecules) at four different densities (0.95, 1.0, 1.1, and 1.2 g/cm$^3$) 
with the empirical force field TIP4P.\cite{TIP4P} These simulations were equilibrated at 300 K along 100 ps runs using a Nose-Hoover thermostat. 
Subsequently, they were allowed to continue for 1 ns,  using a Verlet integrator.\cite{Annealing}

%Jue added 1.0 PBE data point into this graph on Aug 22rd

\begin{figure}[!hbtp]
\begin{center}
\includegraphics*[width=8.0 cm]{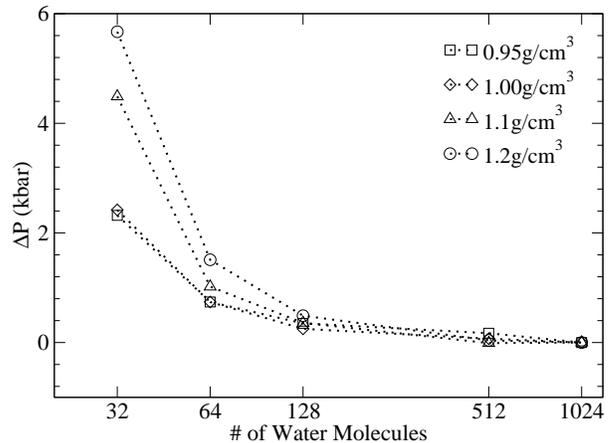}
\caption{ Convergence of average pressure with system size. $\Delta P$ is the pressure difference between the system of $N$ molecules and that of 1024 molecules, at 300 K and four different densities.  
All the simulations in this figure were performed using the TIP4P empirical force field.} 
\label{P_sizeEffect}
\end{center}
\end{figure}

The convergence of average pressure of TIP4P simulations with system size is shown in Fig.~\ref{P_sizeEffect}. 
Simulations with 32 water molecules result in a large discrepancy with the converged value. 
With 64 molecules the pressure is not yet fully converged but the error is much smaller.
Considering the need for long simulation times and the large number of simulations required for this study, we chose to limit our systems to 64 water molecules 
to study the density-pressure dependence.

\section{SIMULATION RESULTS} 
\subsection{Equilibrium density for GGA functionals}

Table \ref{tab:All-simulations} summarizes the parameters used, and the average values obtained, in all our AIMD simulations, spanning five functionals and a wide range of pressures and densities.
As explained before, the simulation temperatures were chosen, for each functional, to achieve a reasonable comparison with the experimental diffusivities and radial distribution functions.\cite{mvfs04,mvfs04-1}

\begin{table*}[!hbtp]
\caption{Simulations parameters used, and average values obtained in this work: 
mass density $\rho$, 
exchange-correlation functional,
average temperature $T_{avg}$, 
diffusion coefficient $D$, 
position $r^{max}_\mathrm{OO}$ of first maximum  in $g_\mathrm{OO}(r)$, 
position $r^{min}_\mathrm{OO}$ of first minimum in $g_\mathrm{OO}(r)$, 
average coordination number, and 
average number of hydrogen bonds $N_\mathrm{H-bond}$. 
For each calculation, quantities are averaged over 20 ps. 
At ambient conditions, the experimental self-diffusion coefficient is $2.2 \times 10^{-5}$ cm$^2$/s 
for H$_2$O and 1.8$\times$10$^{-5}$ cm$^2$/s for D$_2$O\cite{waterexpD}. 
The experimental coordination number is 4.7\cite{RDFexp2}} 
\centering
\resizebox{13cm}{!}{
\begin{tabular}{c c c c c c c c }
\hline
\hline
$\rho$ (g/cm$^3$) & Functional & $T_{avg}$ (K) & $D$ ($10^{-5}$cm$^2$/s) & $r^{max}_\mathrm{OO}$ (\AA) & $r^{min}_\mathrm{OO}$ (\AA) & Coord. & $N_\mathrm{H-bonds}$ \\
\hline
0.65 & BLYP & 361 & 2.82  &2.87 & 3.63 & 3.85 & 3.49\\
0.75 & BLYP & 361 & 3.26  &2.86 & 3.53 & 3.89 & 3.40\\
0.85 & BLYP & 349 & 2.29  &2.83 & 3.45 & 4.03 & 3.58\\
0.95 & BLYP & 349 & 2.01  &2.82 & 3.40 & 4.33 & 3.61\\
0.65 & PBE & 363 & 3.72 & 2.81 & 3.52 & 3.77 & 3.43\\
0.75 & PBE & 358 & 2.49 & 2.80 & 3.47 & 4.21 & 3.51\\
0.85 & PBE & 352 & 1.33 & 2.77 & 3.41 & 4.28 & 3.67\\
0.95 & PBE & 354 & 1.16 & 2.76 & 3.36 & 4.31 & 3.69\\
%Jue modified this table on Aug 23rd.
1.00 & PBE & 353 &  1.18 & 2.75 & 3.28 & 4.23  & 3.78\\
0.65 & revPBE & 338 & 4.35 & 2.89 & 3.62 & 3.58 & 3.34\\
0.75 & revPBE & 339 & 4.19 & 2.88 & 3.55 & 3.79 & 3.56\\
0.85 & revPBE & 348 & 3.73 & 2.85 & 3.51 & 4.09 & 3.46\\
0.95 & revPBE & 341 & 2.74 & 2.83 & 3.45 & 4.46 & 3.50\\ 
0.95 & DRSLL & 296 & 2.68 & 2.93 & 3.50 & 4.87 & 3.63\\
1.00 & DRSLL & 300 & 2.63 & 2.92 & 3.46 & 4.90 & 3.68\\
1.05 & DRSLL & 303 & 2.12 & 2.92 & a & - & -\\
1.00 & DRSLL-PBE & 304 & 2.08 & 2.83 & 3.38 & 4.52 & 3.58\\
\hline
\hline
\end{tabular}
}
\footnotetext[1]{As there is no sharply identifiable first minimum in $g_\mathrm{OO} (r)$ at 1.05 g/cm$^3$, we can not provide the accurate value here.}
\label{tab:All-simulations}
\end{table*}

% Marivi, is it ok not to mention how do we get the error bars in this figure?

\begin{figure}[!hbtp]
\begin{center}
\includegraphics*[width=8.0 cm]{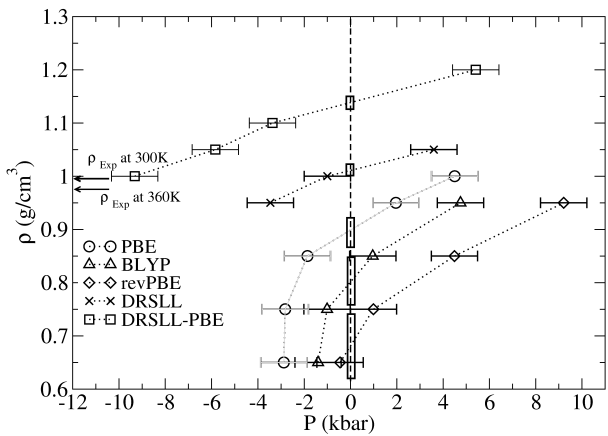}
\caption{Pressure-density curves obtained in AIMD simulations with different GGA and vdW XC functionals. The error bars are standard deviations of average pressures in three different runs with the same density, XC functional, and temperature. The simulation temperatures  can be found in Table \ref{tab:All-simulations}. The colored boxes show the estimated range of equilibrium densities at $P=1$ atm $\simeq 1$ bar. The arrows indicate the experimental densities at 300 K and 360 K and 1 atm. \cite{Rou&T_Exp}}
\label{Fig1}
\end{center}
\end{figure}

Pressure-density curves obtained with three GGA functionals (BLYP, PBE and revPBE) and two vdW-DFs (DRSLL and DRSLL-PBE)  are compared in Fig.~\ref{Fig1}. 
%The simulation temperatures  can be found in Table \ref{tab:All-simulations}. 
%The pressure error bars, which are on the order of 1 kbar, are estimated as standard deviations from three different simulations using the same density, XC functional, and temperature, but different initial configurations. 
Because of the reasons given in the introduction, each functional is simulated at a different temperature (see Table \ref{tab:All-simulations}), but we would like to stress that this has only a minor effect on the resulting average pressures.\cite{Rou&T_Exp,Rou&T_TIP5P} 
To illustrate this, we show in Fig.~\ref{Fig1} the experimental densities at 300 K and 360 K for $P=1$ atm.
Their difference is much smaller than that between different functionals.

All GGA functionals result in theoretical equilibrium densities considerably lower than the experimental value ($\rho=0.997$ g/cm$^3$ at $P=1$ atm). 
The equilibrium density of PBE is 0.85-0.90 g/cm$^3$, close to the value presented in Refs.~\onlinecite{Schwegler00,Schmidt09} and around 12$\%$ lower than experiments. 
The BLYP equilibrium density of 0.76-0.85 g/cm$^3$, also within the range provided by Schmidt {\it et al}\cite{Schmidt09} and 19$\%$ lower than experiments.
The revPBE density is 0.63-0.75 g/cm$^3$, or 31$\%$ less dense than experiments.

The large differences in the calculated densities of similar GGAs are surprising and require some analysis.
Like many liquids, water maintains much of the short-range order of its solid.
In ice Ih, a rigid framework of hydrogen bonds forces a tetrahedral coordination and a relatively open structure with large voids.
Four of these interstitial voids surround each molecule, along orientations opposite to those of its H bonds (we will call them `anti-tetrahedral' orientations).
In the liquid, entropy implies that part of the H bonds are broken and part of the interstitial voids are occupied.
Thus, we can imagine three different properties of a functional or force field, that will determine the density of liquid water:
\textit{i}) the length of the hydrogen bonds, causing a possible dilation of the H-bond framework; 
\textit{ii}) the strength of these bonds, that determines the average H-bond coordination; and 
\textit{iii}) the potential energy of molecules at interstitial positions, that determines their average occupation.

With a rigid H-bond framework, the density would be proportional to $d_\mathrm{OO}^{-3}$, where $d_\mathrm{OO}$ is the average O-O distance. 
The change in $d_\mathrm{OO}$
between PBE and revPBE (see Table~\ref{tab:All-simulations}) is $\sim$3\%, which would
translate into a difference in density of just $\sim$10\%, far from the $\sim$25 \% observed.

However the density also depends on the number of H bonds, which increases with their strength. 
Fig.~\ref{Antidimer}(b) shows that, despite the similarity of the two functionals, the H bond energy is $\sim$50\% larger with PBE than with revPBE.
As a consequence, the first coordination peak in the O-O radial distribution function (RDF) $g_\mathrm{OO}(r)$, plotted in Fig.~\ref{Fig2}, is $\sim$30\% higher with PBE, showing that the H-bond network is better preserved with this functional.
%Fig.~\ref{HBcoord1} also shows that there are more H bonds per molecule in PBE-water than in revPBE water. 
All put together, the difference in densities within the first coordination shell, calculated as $\frac{N_{coor}+1}{4\pi r^{min}_\mathrm{OO}/3}$ (with values from Table~\ref{tab:All-simulations}), where $N_{coor}$ is the average coordination number and $r^{min}_\mathrm{OO}$ is the radius of the first minimum in $g_\mathrm{OO}(r)$, is larger than 30$\%$.
%Looking at the data in the region of $\rho=1$ g/cm$^3$ and for positive pressures, 
%Also, the isothermal compressibility, 
%$\beta =\frac{1}{\rho} \frac{\partial \rho }{\partial P}$, is slightly larger in our simulations for revPBE than with PBE, reflecting the weaker H-bond %network produced by this functional.

% Jue modified this section on Aug30, 2010
% Marivi, Shall we modified the name of this section?
\subsection{GGA functionals: effect of pressure on the RDF}
\label{sec:GGARDF}

\begin{figure}[h!btp]
\begin{center}
\includegraphics[width=8.0 cm]{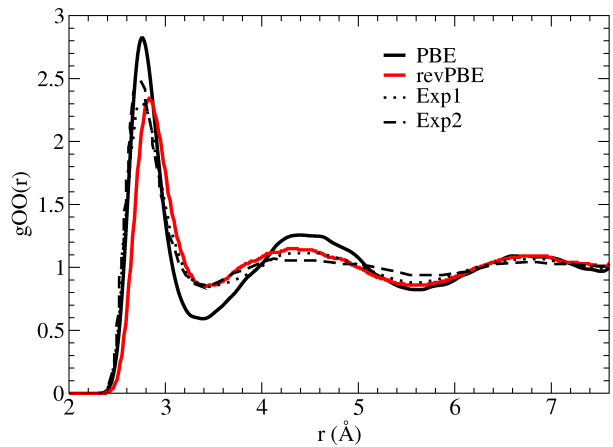}
\end{center}
\caption{O-O radial distribution functions (RDF) of liquid water for AIMDs in this work, at 0.95 g/cm$^{3}$ with PBE at 360 K (solid black line)  with revPBE at 340 K (solid red line). These temperatures are chosen\cite{mvfs04} so that the structure and diffusivity compare best with those of experiments at 300 K. Exp1 (dotted line) and Exp2 (dashed line) data are the experimental data in Refs.~\onlinecite{RDFexp3} and \onlinecite{RDFexp4} respectively.}
\label{Fig2}
\end{figure}

Fig.~\ref{Fig2} compares the O-O RDF of liquid water at 0.95 g/cm$^3$ for the PBE and revPBE functionals. 
   The temperature difference between the the two simulations is under 15 K, and should have a negligible effect on the RDFs\cite{Schwegler00} compared with that resulting from pressure differences. 
We compare our results with two different experimental data sets\cite{RDFexp3,RDFexp4}, both obtained from Ref.~\onlinecite{RDFexp4}. 
The revPBE RDF matches the experiments rather well. Although the 
onset and position of the first peak are clearly shifted to the right,
there is a large uncertainty in the experimental data for this peak
(see Ref.~\onlinecite{RDFexp4}). However, at 0.95 g/cm$^3$, revPBE water is at a very
large pressure. When this pressure is released, the density drops to 
$\sim$0.65 g/cm$^3$ and the agreement with the experimental RDF is
totally lost (see Supplementary Information). Both the first and
second peaks move further to the right.
This effect is also observed in PBE.
It should be noted that, for PBE, the height of the first minimum remains well below the experimental value and is barely modified, either by temperature or pressure effects (as seen in supplementary information).

\subsection{Van der Waals Interactions in Water}
\label{sec:DRSLL}

Van der Waals (vdW) dispersion interactions, due to non local electron correlations, are not treated properly by the local density\cite{Ceperley80} (LDA) and GGA\cite{metagga1,PBE,XWu01} functionals, in which electron correlations are treated as local or semi-local effects.
It is known that in water, due to the high polarizability of oxygen,\cite{tsiper05,tsuzuki01,XWu01}
vdW interactions have a significant contribution to the binding.
The vdW attraction contributes to strengthening both H-bonds and non-H-bond interactions\cite{Chunlin09} and it increases the overall cohesive energy in the liquid. 
Schmidt {\it et al} have studied the vdW effect on the density of water with the PBE-D method,\cite{Grimme} which includes an interatomic pair potential correction added to the PBE functional.\cite{Schmidt09} 
They showed that the density of PBE-D water is very close to the experimental value, and the resulting liquid is also
structurally closer to experiments. The dynamical properties of their liquid were not accessible due to the use of a temperature and pressure thermostat
in their simulations.

Furthermore, classical force fields, that represent electronic dispersion
 correlations with $-1/r^6$ interatomic pair potentials, can be inaccurate 
 because these correlations are not atom-centered in principle and they 
 depend on the instantaneous environment of each atom and molecule.

Since the current DFT description of liquid water overestimates intermolecular binding, it is not clear whether vdW interactions will reduce the difference with experiments of various magnitudes. 
However, the attractive vdW effects could be especially relevant to increase the density of the simulated liquid, bringing it closer to experiment, as
Schmidt {\it et al} have seen using the PBE-D method.

 As seen in Fig.~\ref{Fig1}, the equilibrium density computed with the DRSLL vdW-DF is 1.02 g/cm$^3$, only 2\% \textit{above} the experimental value. 
 This is much better than any of the commonly used GGA functionals. 
% The shape of $\rho(P)$ is also different, and DRSLL water has a much smaller compressibility (larger bulk modulus), at $P=1$ atm, than any of the GGAs, %indicative of a stronger H-bond network.

The O-O RDF of DRSLL water show large disagreements with experiments, as shown in Fig.~\ref{Fig4}. O-H and H-H RDFs are presented in the supplementary information. 
The average first neighbor distance, is longer than the experiments.
The second coordination peak in $g_\mathrm{OO}(r)$ `collapses' significantly with this functional, and an anomalous hump at $r=3.8$ \AA\ appears.

Differences in the liquid H-bond interactions can be characterized by relaxing typical liquid-phase structures.\cite{Giovambattista02}
Table \ref{tab:HB} shows several characteristic magnitudes calculated after relaxing five pentamer clusters, randomly chosen from our AIMD,\cite{FootnoteonBasis} with different functionals (but with the same initial geometries for all functionals).
The table shows that DRSLL increases the H-bond binding energy of revPBE but it also yields a longer H-bond, producing the shift of the first coordination peak to longer distances.
This overestimation of DRSLL binding distances has been observed in many other systems.\cite{DRSLL,LangrethJPCM} 

\begin{table}[h]
\centering
\caption{Comparison of average H-bond energy $E_{HB}$, and distance $d_{HB}$, first neighbor O-O distance $d_\mathrm{OO}$, and intramolecular O-H bond length $d_\mathrm{OH}$ obtained by relaxing the same five pentamer clusters with different XC functionals. Each cluster contains a randomly-chosen central molecule plus its first neighbors.}
\begin{tabular}{c c c c c }
\hline
\hline
Functional & $E_{HB}$ (eV) & $d_{HB}$ (\AA) & $d_\mathrm{OO}$ (\AA) & $d_\mathrm{OH}$ (\AA)\\
\hline
BLYP  & 0.222 & 1.921 & 2.881 & 0.978  \\
PBE  & 0.262 & 1.872 & 2.809 & 0.979  \\
revPBE & 0.199 & 2.022 & 2.974 & 0.974  \\
DRSLL  & 0.225 & 2.022 & 2.980 & 0.974 \\
DRSLL-PBE & 0.290 & 1.918 & 2.833 & 0.976\\
\hline
\hline
\end{tabular}
\label{tab:HB}
\end{table}

\begin{figure}
\begin{center}
\includegraphics*[width=8.0 cm]{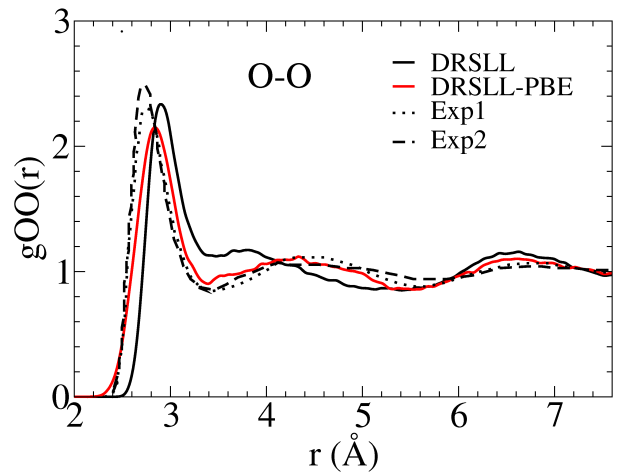}
\caption{Comparison of the O-O RDFs obtained with the DRSLL (black solid line) and DRSLL-PBE (red solid line) functionals, and experimental results of Refs.~\onlinecite{RDFexp3} (dotted line) and \onlinecite{RDFexp4} (dashed line). Both simulations and experiments are at 1.0 g/cm$^3$ and $\sim$300 K.}
\label{Fig4}
\end{center}
\end{figure}

\subsection{Effect of PBE exchange in vdW-DF}

As mentioned before, the exchange energy in the DRSLL functional is taken from the revPBE formulation.\cite{DRSLL}
This choice reduces, relative to PBE, the spurious exchange-induced binding in simple systems like noble gas dimers.\cite{DRSLL}
However, revPBE underestimates the H-bond interaction, what translates to larger O-O distances and a clear under-structuring of the RDF, compared to experiments. 
Therefore, it is worth studying the effect of combining vdW correlations, as described by the DRSLL functional, with a stronger H-bond interaction, as obtained using PBE exchange. 
Thus, we have replaced the revPBE exchange in DRSLL by PBE exchange and we will refer to this functional as DRSLL-PBE. 
Recent studies\cite{Klimes2010,LMKLL} have shown that minor modifications of the exchange enhancement factor, in the same line as that used here, can improve geometries and binding energies of small molecular systems. 
However in this study we chose the original DRSLL and DRSLL-PBE to allow for
comparisons with their GGA counterparts.

We have performed simulations with DRSLL-PBE for 1.00, 1.05, 1.10 and 1.20 g/cm$^3$ at 300 K. 
The data points in Fig.~\ref{Fig1} for this functional represent a single simulation each, and we used the same error bars obtained for DRSLL. 
 At 1.00 g/cm$^3$ average pressures are $\sim$10 kbar, indicating that the strong binding of PBE exchange, added to that induced by vdW correlations,
 has a net effect of over correcting the positive pressure of GGAs. We find
an equilibrium density at 1 atm of 1.13 g/cm$^3$
Indeed, the pressure reduction achieved by DRSLL versus revPBE is 13 kbar, close to that of DRSLL-PBE versus PBE, 14 kbar.

As shown in Fig.~\ref{Fig4}, DRSLL-PBE recovers an O-O RDF similar to experiments, without an artificial increase in the simulation temperature. 
The height of the first minimum in $g_\mathrm{OO} (r)$ is very close to the experimental one, and the anomalous hump displayed by DRSLL is almost completely eliminated.
The fact that the long range tail of $g_\mathrm{OO} (r)$ matches closely the experimental one may be indicative of a correct characterization of the long range structure of water, obtained when vdW interactions are included. We provide a comparison of the structure factors calculated for PBE and
DRSLL-PBE in the supplementary information.

\subsection{Self Diffusion Properties}
\label{sec:Diffusion}

\begin{figure}[!hbtp]
\begin{center}
\includegraphics*[width=8.0 cm]{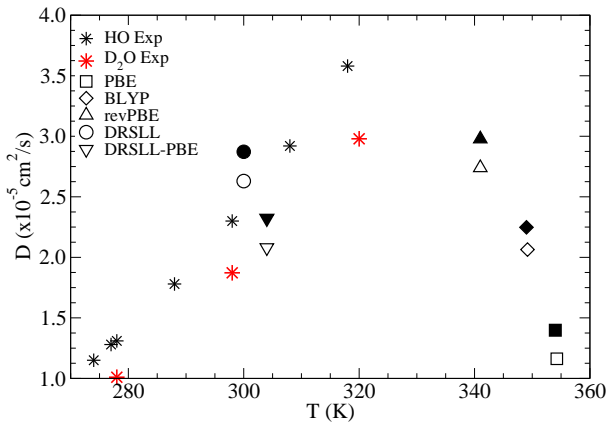}
\caption{Self-diffusion coefficient vs temperature obtained in AIMD simulations with various XC functionals compared to H$_2$O and
D$_2$O experimental values.\cite{waterexpD} Values corrected for finite size effects\cite{Dunweg93} are shown as solid symbols, and observed values  as open symbols. Note that our simulations are performed for D$_2$O.} 
\label{Figdiffusivity}
\end{center}
\end{figure}

In Fig.~\ref{Figdiffusivity}, the temperature dependence of diffusivity, computed from all our AIMD simulations, is compared with experimental values at similar and lower temperatures. 
We also show values corrected for finite size effects, as proposed in Ref.~\onlinecite{Dunweg93}.
 Confirming previous results,\cite{mvfs04} we find that the temperature of AIMD with PBE and BLYP must be 16-20\% higher than the experimental one, to give a similar diffusivity. 
 With revPBE, the temperature needs to be only 6-9\% higher. 

On the other hand, both DRSLL and DRSLL-PBE nearly reproduce the experimental diffusivity at room temperature, without any temperature rescaling. This is not unexpected, given the strong link between structure and dynamical properties of liquid water.\cite{mvfs04} Thus, one of our most important conclusions is that DRSLL-PBE is a very good XC functional to describe liquid water: PBE requires scaling up the temperature to reach an agreement with experiments. DRSLL replicates diffusivity well, but it substantially fails to describe the RDFs. DRSLL-PBE corrects the structure while maintaining a very good diffusivity.

\section{ANALYSIS AND DISCUSSION}

\subsection{Intermolecular interactions}

\begin{figure}[!hbtp]
\begin{center}
\includegraphics*[width=8.0 cm]{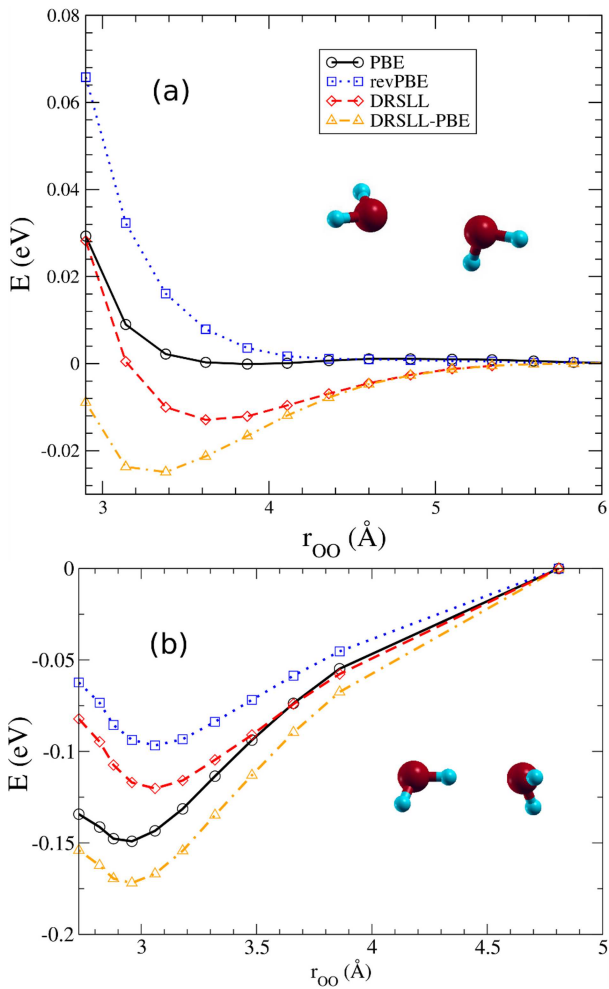}
\caption{Total energy of the water dimer as a function of the intermolecular separation for two different
molecular orientations calculated for PBE (circles), revPBE (squares), DRSLL (diamonds) and DRSLL-PBE (triangles).
(a): Non H-bonded configuration as shown in the inset (with partially facing O lone pairs from each molecule).
 (b): H-bonded configuration as shown in the inset. In both graphs the energies have been shifted to have the zero
at the largest separation.} 
\label{Antidimer}
\end{center}
\end{figure}

Since the vdW attraction of vdW-DFs does not reduce the H-bond distances (actually they increase, relative to their GGA counterparts), what causes the large increase in equilibrium density?
One possibility is that the stronger H-bonds, shown in Fig.~\ref{Antidimer}(b), increase the average first-neighbor coordination.
However, the height of the first peak in $g_\mathrm{OO}(r)$, as shown in Figs.~\ref{Fig2} and \ref{Fig4}, actually decreases from the GGAs to the vdW-DFs. 
Therefore, we are left only with the third possibility mentioned before, i.\ e.\ an increase in the occupation of the `interstitial', non-H-bonded sites.

In order to characterize the vdW binding between water molecules, we have calculated the interaction energy between two water molecules oriented as shown in Fig.~\ref{Antidimer}(a), thus avoiding the hydrogen bond interaction.  
By comparison, we also use PBE and revPBE. 
Fig.~\ref{Antidimer}(a) clearly shows that DRSLL exhibits a minimum in the potential at $\sim$3.7 \AA\ with a binding energy of 10 meV, not shown by the GGAs. 
Although this is many times times weaker than the H-bond, it will have an important effect in increasing the occupation of the interstitial sites, which are roughly at that distance.
It is worth noting that the position of this potential energy minimum is close to the first minimum of 
$g_\mathrm{OO} (r)$, obtained with the GGAs.
This means that this new `vdW bond', and its effect on increasing the occupation of interstitial sites, may account for  the unusual hump appearing in the $g_\mathrm{OO} (r)$ of DRSLL at this distance, as seen in Fig.~\ref{Fig4}. 
The potential energy minimum in DRSLL-PBE is even deeper, 25 meV, and shifted to shorter distances (3.4 \AA). This is also the origin of the increase of the height of the first $g_\mathrm{OO} (r)$ minimum from PBE to DRSLL-PBE.

We have also computed the H-bond potential energy curve between two water molecules in a H-bound configuration. The results are shown in Fig.~\ref{Antidimer}(b).
The depth of the potential is lower that the optimal H-bond interaction energy because the geometry of the molecules was not optimized. 
Still the results show the same tendency observed in the RDF and in Table ~\ref{tab:HB}. 
When the vdW functional is included, the energy of the H-bond interaction increases by approximately 25 meV  independently of the GGA used to describe the exchange interaction.
Therefore, the weakening of the H-bond network by vdW interactions (reduction of the first and second coordination peaks) is not associated to the weakening of the H-bond itself, but to the increase of new, favorable, non H-bonded configurations that compete with the H-bonds. 

% Jue Modify the Fig5 (top) with Marivi's new graph.

%To explore further this relationship, we show in Fig.~\ref{rho-difference} the difference between GGA and vdW RDFs, multiplied by their mean densities.
%This difference gives the excess density, as seen from a molecular center. 
%Three features are relevant in this graph.
%First we observe a large, negative peak at $\sim 2.7$ \AA\ . 
%This is followed by a maximum at $\sim 3.1$ \AA\ . 
%These two features are due to the right shift of the first coordination peak  in vdW-DFs, with respect to the corresponding GGAs. 
%Third, we also observe a hump at $\sim 3.4$ \AA\ for DRSLL-PBE and at $\sim 3.8$ \AA\ for DRSLL, that occurs exactly at the positions of the vdW potential minima shown in Fig.~\ref{Antidimer}(a) and marked by arrows in Fig.~\ref{rho-difference}.

%\begin{figure}[!hbtp]
%\begin{center}
%\includegraphics*[width=8.0 cm]{deltarho.eps}
%\caption{Excess density (see definition in text) relative to a molecular centre, for DRSLL ($\rho=1.0$ g/cm$^3$) compared to revPBE ($\rho=0.95$ g/cm$^3$), red dotted lines and DRSLL-PBE ($\rho=1.0$ g/cm$^3$) compared to PBE ($\rho=0.95$ g/cm$^3$), black line. 
%The arrows indicate the position of the energy minima in Fig.~\ref{Antidimer}(a).} 
%\label{rho-difference}
%\end{center}
%\end{figure}

\subsection{Spatial Distribution Functions}

\begin{figure*}[!hbtp]
\begin{center}
\includegraphics*[width=17.0 cm]{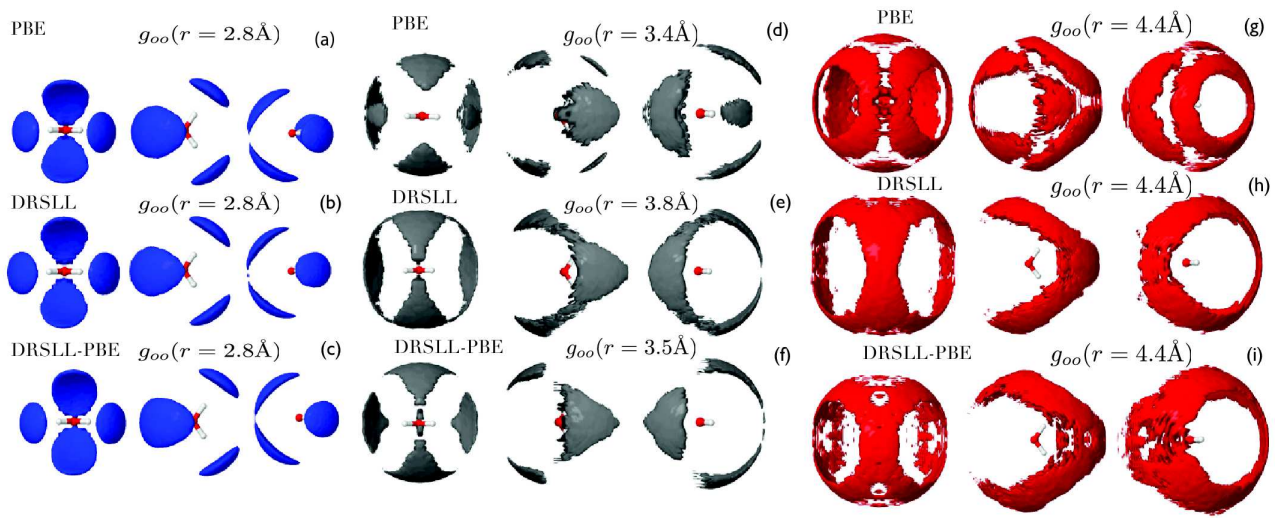}
\caption{Spatial distribution function(SDF) of liquid water, from 3 viewpoints (front,top and side) and for 3 different
shell radii as indicated in the legend: (a) PBE, first coordination shell, (b) DRSLL, first coordination shell, (c) DRSLL-PBE, first coordination shell,  (d) PBE, interstitial shell, (e) DRSLL, interstitial shell, (f) DRSLL-PBE, interstitial shell, (g) PBE, second coordination shell, (h) DRSLL, second coordination shell, and (i) DRSLL-PBE, second coordination shell.}
\label{SDF}
\end{center}
\end{figure*}

RDFs give angular-integrated information, but the angular distribution of molecules around a given one also contains very valuable information that can differentiate between similar RDFs.
In Fig.~\ref{SDF} we plot the O-O spatial distributions functions (SDF)\cite{svishchev93} $g_\mathrm{OO} (r,\theta,\phi)$ for three functionals. 
The polar angles $\theta,\phi$ are referred to a local coordinate set of the central molecule, with origin at the oxygen atom: $x$ (direction of the $\widehat{\mathrm{HOH}}$ angle bisector), $y$ (perpendicular to $x$, within the molecular plane) and $z$ (normal to the molecular plane).
We plot isosurfaces $g_\mathrm{OO}(r,\theta,\phi)=g_c$, restricted to spherical shells of thickness $\delta r=0.2$ \AA\  centered at three different distances $r$: 
\textit{i}) at the first maximum of $g_\mathrm{OO} (r)$, with $g_c=2$, Figs.~\ref{SDF}(a-c); 
\textit{ii}) at the first minimum of $g_\mathrm{OO} (r)$, with $g_c=0.5$, Figs.~\ref{SDF}(d-f); and 
\textit{iii}) at the second maximum of $g_\mathrm{OO} (r)$, with $g_c=1$, Figs.~\ref{SDF}(g-i). 
To ease the visualization we show three different viewpoints (front, top and side) for each shell.

Fig.~\ref{SDF} compares the O-O SDF of PBE, DRSLL, and DRSLL-PBE water at 1.0 g/cm$^3$. We omit revPBE because of its close similarity to PBE. 
The structure of the first maximum is almost identical for all the functionals considered, i.e, nearly tetrahedral. 
The density of acceptor molecules (lobes in front of the H atoms) is much more localized than the density of donor molecules (lobes in the region of the oxygen lone pairs behind the oxygen atom) which is more disperse. This dispersion reflects the asymmetry of the H bonds, which are less directional on the accepting molecule. 

The most important differences between the vdW and GGA functionals occur beyond this tetrahedral shell, at the distance of the first minimum of $g_\mathrm{OO} (r)$ (Figs.~\ref{SDF}(d-f)).
%This is the region marked by the arrows in Fig.~\ref{rho-difference}, that corresponds also to the minima of the vdW potential (Fig.~\ref{Antidimer}(a)). 
At this distance, the angular distribution for the PBE functional has small lobes in the tetrahedral directions and larger ones in the opposite directions, that we will refer as anti-tetrahedral.\cite{SoperPRL2000}
The small tetrahedral lobes correspond to the tail of the first coordination peak. 
The anti-tetrahedral lobes correspond to the directions of the interstitial sites, and it is fully dominant in the vdW-DFs (Figs.~\ref{SDF}(e-f)).
Thus, although this anti-tetrahedral shell is already present with the GGA, due to entropic effects, its density largely increases in the vdW-DF due to the vdW attraction. 
Soper et~al.\cite{SoperPRL2000} observed a similar effect in liquid water under pressure. 
Thus, the effect of non local vdW correlations in DFT water is similar to the effect of pressure in experimental water.

The third shell shown in Figs.~\ref{SDF}(g-i), corresponds to the position of the second coordination shell, at $r \sim 4.4$ \AA. %and to the second minimum in Fig.~\ref{rho-difference}. 
This peak, formed by second neighbors, tetrahedrally coordinated to the first coordination shell, is depleted in vdW water in favor of the interstitial anti-tetrahedral shell.

We can conclude from this analysis that the effect of vdW-DF correlation is to increase the density of the liquid by populating the interstitial sites that are at a distance between the first and second coordination shells. These interstitial structures become more favorable because of non-H-bonded vdW interactions, as shown in Fig.~\ref{Antidimer}(a), and they compete with, and destabilize, the H-bond network.
They also increase the height of the first minimum in $g_\mathrm{OO} (r)$, that is too low with PBE.
On the other hand, revPBE water is already relatively unstructured, H-bonds being rather weak within this functional, as shown in Table \ref{tab:HB}. 
Therefore the density increase induced by vdW interactions, combined with a weak H-bond network, results in a collapse of the structure, characteristic of water under pressure.

\subsection{Hydrogen Bond Network}

In previous sections we did not explicitly analyze the hydrogen bond network.
In water, each molecule tends to be surrounded by four others, donating two H-bonds and receiving two others in a nearly tetrahedral arrangement. This translates into a H-bond network with a majority of fourfold-coordinated molecules.
 However it also contains under-coordinated and over-coordinated molecules, and the population of under-coordinated molecules correlates with an increasing self diffusion constant. The average bond life-time of the H-bonds in liquid water is of the order of 1 ps. \cite{HBlifetime}

%% New part, we show correlations and anticorrelations of Hb-VdW bonds

One of the most striking differences between the GGA functionals and their vdW-DF counterparts is the large change observed in the first
peak of the O-O RDF. In average, this implies both a reduced number of H-bonds and longer H-bond lengths when vdW interactions
are accounted for. This however does not correlate with the average H-bond energies and geometries presented in table \ref{tab:HB}. The table shows that vdW interactions increase the H-bond energy by $\sim$25 meV, both for revPBE and PBE. This is close to the vdW (non-H-bond) binding energy shown in Fig.~\ref{Antidimer}(a). Also, the right shift of the O-O RDF peak (2.5$\%$) is much larger
than the increase in average O-O distances shown in table \ref{tab:HB} (0.9$\%$).  
To study the origin of these deviations we have analyzed the statistics and geometries of:
(i) H bonds, defined by the molecules that contribute to the first, tetrahedral, shell shown in Fig.~\ref{SDF}(a-c) and (ii) VdW bonds,
or interstitial molecules in anti-tetrahedral coordination to the H-bond network as shown in Fig.~\ref{SDF}(d-f). 
The total $g_\mathrm{OO} (r)$ has been divided in three contributions: first-shell and second shells of H bonds (HB1 and HB2) and a reminder $g^{total}_\mathrm{OO} (r)-
g^{HB1}_\mathrm{OO} (r)-g^{HB2}_\mathrm{OO}(r)$. 

Results are shown in Fig.~\ref{fig:rdfdecomposition} and 
 Table~\ref{tab:H-bonds-vdwbonds} for PBE and DRSLL-PBE, 
 and in Supplementary Information for revPBE and DRSLL.
VdW correlations produce slightly less H-bonds per molecule (3.78 vs 3.81) with a broader distribution of lengths.
The second shell of H-bonds, represents $\sim 60\%$ of the second peak in the RDF. 
Again, there are slightly less second-H-bonded molecules with DRSLL-PBE (10.47 vs 10.59). 
These numbers are very close to the ideal `mean-field' value $N_{HB2}=N_{HB1}(N_{HB1}-1)$. The bottom plot in Fig.~\ref{fig:rdfdecomposition} represents all the neighbors not included in the previous two distributions. These include the third shell of H-bonds and the vdW-bonded molecules (these two are not mutually exclusive). We see an important 20$\%$ increase in DRSLL-PBE, up to a distance of 4.5 \AA, a region in which non-H-bonded interstitial molecules are dominant. 
This increase is about twice as large from revPBE to DRSLL (see Supplementary Information).
%ANADIR ESTO EN LA SI
Notice that the maximum of this distribution for DRSLL-PBE, at 3.5 \AA, coincides approximately with the position of the vdW energy minimum of Fig.~\ref{Antidimer}, and also with the first minimum of the total $g_\mathrm{OO}(r)$. 
We have also observed a positive correlation between the number of non-H-bonded neighbors of a molecule and its average H-bond distances. 
This produces a 1.7$\%$ right shift of the average O-O distance for the first H-bonded shell from PBE to DRSLL-PBE, due to the increase of vdW bonds.
A more detailed analysis of this, and other related effects, will presented in a future contribution.

\begin{figure}[!hbtp]
\begin{center}
\includegraphics*[width=8.0 cm]{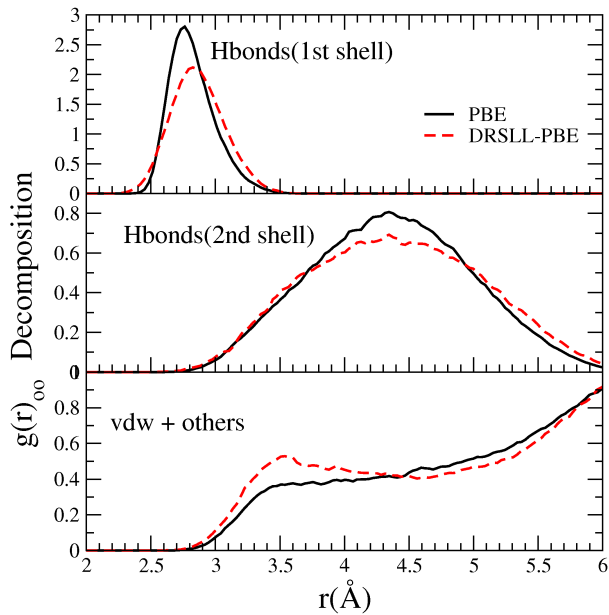}
\caption{Decomposition of $g_\mathrm{OO} (r)$ into H and vdW bonds. 
%Solid black lines are PBE and dashed red lines are DRSLL-PBE results. 
Top, tetrahedral shell of first H-bonded neighbors, shown in Fig.~\ref{SDF}(a-c). Center, second H-bonded shell shown in Fig.~\ref{SDF}(g-h). Bottom, remaining molecules which include vdW-bonded, interstitial and third H-bonded molecules.
The maximum of the lower DRSLL-PBE curve, at $r$=3.5\AA, is due to the molecules in anti-tetrahedral coordination to the H-bond network, as shown in Fig.~\ref{SDF}(d-f).
Simulations in this figure are for liquid water at 1.0 g/cm$^3$ and $\sim$300 K.}
\label{fig:rdfdecomposition}
\end{center}
\end{figure}

\begin{table}[h]
\centering
\caption{Mean values for H and vdW bonds at 1.0 g/cm$^3$ and $\sim$300 K. $N_{HB1}$ and $N_{HB2}$: average number of H-bonded molecules in the first and second coordination shells (integral of first two curves in Fig.~\ref{fig:rdfdecomposition}). $N_{vdW}$: average number of non-H-bonded molecules (vdW-bonded and others) in a sphere of 4.5 \AA\ radius. $r_{HB1}$: average O-O distance for first H-bonded shell.}
\begin{tabular}{c c c c c c c}
\hline
\hline
Functional & $N_{HB1}$ & $N_{HB2}$ & $N_{vdW}$ & $r_{HB1}$ (\AA)   \\
\hline
PBE       & 3.81 & 10.59 & 3.25 & 2.847 \\
DRSLL-PBE & 3.78 & 10.47 & 3.89 & 2.895 \\
\hline

\hline
\hline
\end{tabular}
\label{tab:H-bonds-vdwbonds}
\end{table}

\section{CONCLUSIONS}

We have performed DFT-based AIMD simulations of liquid water with different GGA and vdW functionals and different densities, comparing the structural and diffusive properties.
The main conclusions are:
\begin{itemize}
\item{} Liquid water under standard atmospheric pressure, as obtained  using standard GGAs, 
is much less dense than experiments: PBE 12\% less, BLYP 19\% less and revPBE 31\% less.
Accordingly, the average calculated pressure at the experimental density is as high as 1 GPa.
\item{} The net effect of vdW non-local correlations is to reduce the pressure and to increase the equilibrium density, which is just 2\% larger than experiments with DRSLL and 13\% higher with DRSLL-PBE.
\item{} With vdW interactions, non-H-bonded water-water configurations are introduced, due to the occupation of interstitial sites. 
These configurations have an anti-tetrahedral orientation, opposite to H-bonds, and they contribute to increase the height of the first minimum in  $g_\mathrm{OO}(r)$. 
While the interstitial sites are partially occupied due to entropy in GGA water, 
their population increases greatly when the vdW attraction is added.
The use of angular-resolved spatial distribution functions is necessary to understand the overall structural changes.
\item{}These anti-tetrahedral structures are the key factor to increase the self diffusion of AIMD water. The diffusivity obtained with DRSLL is very close to experiment at room temperature, without the need of temperature rescaling. 
 \item{}  DRSLL water is under-structured when compared to experiments, with a significant collapse of the second coordination shell induced by the increased density. This is related to the use of revPBE as local exchange, which produces too weak H-bonds. The combination of non local vdW correlation, as in DRSLL, combined with PBE exchange, produces a better liquid, with structural and dynamical properties not far from experiments, without the need of any temperature rescaling. However PBE H-bonds are too strong and DRSLL-PBE causes an over correction of the density with respect to PBE.
\item{} A better treatment of the exchange interaction should improve the H-bond description and, together with vdW correlations, might end up providing the final overall agreement with experimental results in terms of structure, density, and diffusivity of liquid water. A modification of the vdW density functional has been proposed very recently,\cite{LMKLL} and a study of this and other flavors of vdW, not considered here, are under study and will be published in a future work.

\end{itemize}
\section{ACKNOWLEDGEMENT}
We thank Erica L. Lai for her classical MD simulations. The work at Stony Brook University was supported by DOE award numbers DE-FG02-09ER16052 and
 DE-FG02-08ER46550 . The one in Madrid, by MCI grant FIS2009-12712. This research utilized resources at the New York Center for Computational Sciences at Stony Brook University/Brookhaven National Laboratory which is supported by the U.S. Department of Energy under Contract No. DE-AC02-98CH10886 and by the State of New York.


\begin{thebibliography}{63}
\bibitem[{Foo()}]{FootnoteonBasis}
\bibinfo{note}{All the pentamer calculations in this table are performed using
  the full basis for the 64 molecules within the unit cell, as taken from the
  snapshot, and therefore are BSSE (basis set superposition error) corrected.}
\end{thebibliography}
\end{document}